\documentclass[twocolumn,twoside]{revtex4}
\usepackage{graphicx}
\usepackage{fancyhdr}
\begin{document}

\title{Unanswered Questions in Charmed Baryon Physics}

%

\author{John~M.~Yelton}
\affiliation{University of Florida, Gainesville, FL, USA, 32611}

\begin{abstract}
This is an experimentalist's list of questions concerning the physics of the charmed baryon
sector which have no satisfactory answer.
\end{abstract}

\maketitle

\thispagestyle{fancy}


\section{Introduction}

In many conferences experimentalists talk about carefully justified results rather
than analyses they are actively pursuing, and theorists often present work on models which
pertain to measurements that cannot be performed in the near future. Sometimes this leads to 
a disappointing lack of real communication between the two groups. 
I give this talk as an experimentalist, pointing out features of the data
that I do not believe have had satisfactory explanations, but which might in fact be 
easily comprehensible. All the data presented is freely available on the web and I am not
representing any experiment.

\section{The $\Lambda_c(2880)^+$}

In $\Lambda_c^+$ spectroscopy many states have been found, and can all be explained
from a model where the $\Lambda_c^+$ comprises a heavy (charm) quark and a light di-quark.
This configuration predicts a plethora of states. For instance, the addition of 
one unit of orbital angular
monentum can be placed between  the heavy quark and light di-quark (a ``$\lambda$'' excitation) or between
the light quarks in the di-quark (a ``$\rho$'' excitation). Once two units of orbital angular momentum are allowed, 
there are 3 different possiblities of their placement. 
Each of these combinations, with its particular ``light-quark degrees of freedoms'', 
can then can combine with the heavy-quark spin to make 
doublets. 

\begin{figure}[h]
\centering
\includegraphics[width=80mm]{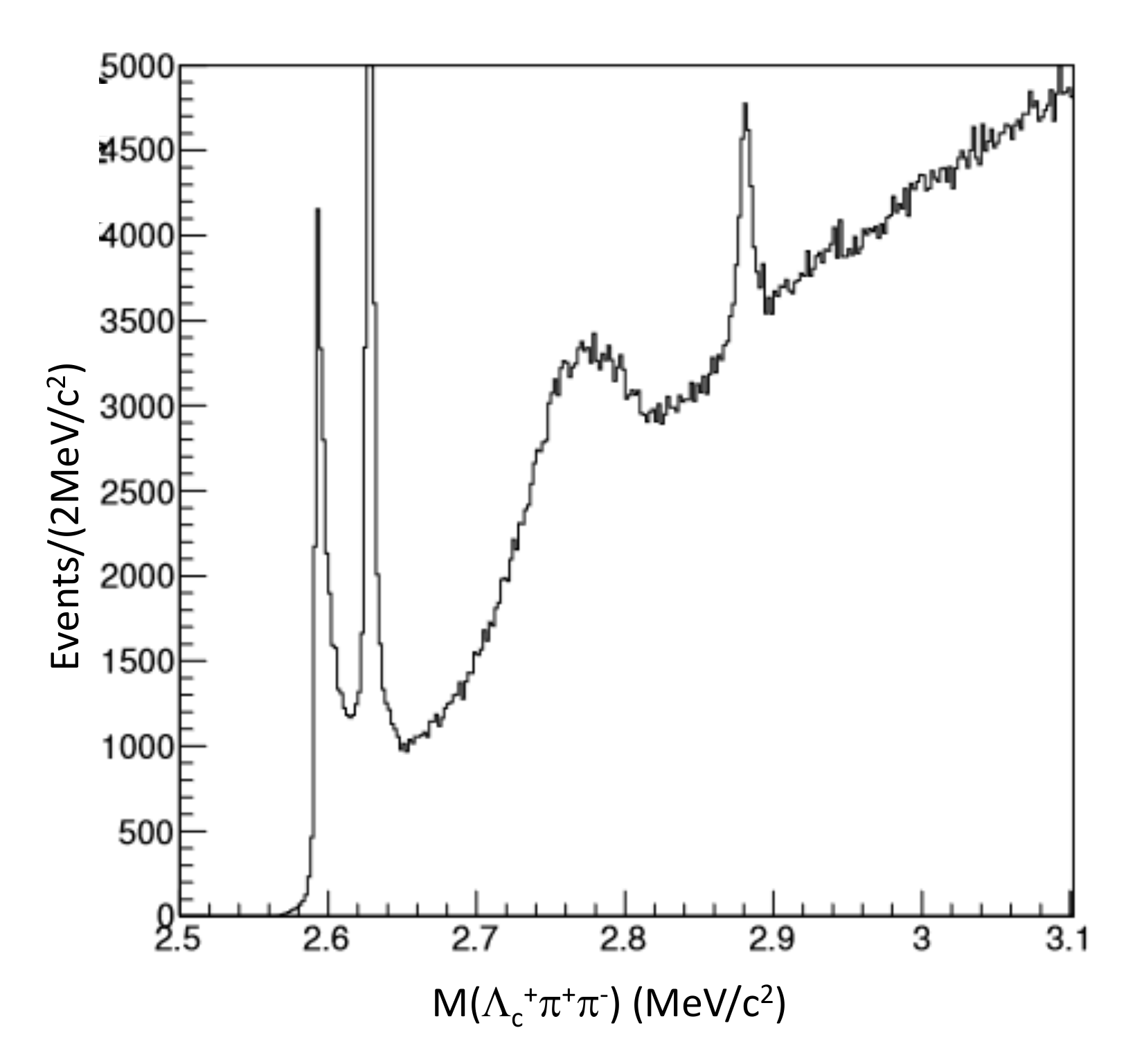}
\caption{The mass difference, $M(\Lambda_c^+\pi^+\pi^-)-M(\Lambda_c^+)$, Belle data. There are four clear
signals, from right-to-left, the 
$\Lambda_c(2880)^+$, 
$\Lambda_c(2765)^+$, 
$\Lambda_c(2625)^+$, and 
$\Lambda_c(2593)^+$. 
} 
\label{fig:LCpp}
\end{figure}

Figure~\ref{fig:LCpp} shows the M($\Lambda_c^+\pi^+\pi^-$) mass
in $e^+e^-$ annihilation data from Belle~\cite{Thesis}. 
The $\Lambda_c(2880)$ sticks out from the background even though there must be many states 
around that mass - this feature caused a surprise when first found~\cite{CLEO2880}.
The state is generally considered to have spin-parity of   $J^P = \frac{5}{2}^+$
based on measurements by Belle~\cite{B2880} and LHCb~\cite{LHC2880}. The natural explanation is
that this is a $\lambda = 2$ excitation. 
My question is, why is this one state so narrow? Data from LHCb~\cite{LHC2880} implies that  
its $J^P\ =\ \frac{3}{2}^+$ partner
is 20 ${\rm MeV}/c^2$ lower and is too wide to be seen clearly in a plot such as this.
In addition,  
the $\Lambda_c(2880)^+$ state has a higher branching fraction into $\Sigma_c(2455)\pi$ than into 
$\Sigma_c(2520)\pi$, whereas the
rules of decay for particles of particular spin-parities, the latter should be preferred as it can 
occur as a  ``P-wave'', whereas the former is restricted to ``F-wave''. It is as if the matrix 
element for the former tends towards zero and suppresses the decay to $\Sigma_c(2520)\pi$
so that the state is narrow. Why is this?

\section{The $\Lambda_c(2593)$}

As we go down to lower masses in the same spectrum (Fig.~\ref{fig:LCpp}), there is a very large $\Lambda_c(2765)$ 
resonance~\cite{LC2765} which 
is only given one ``star'' in the Particle Data Book~\cite{PDG} but clearly exists, and is generally 
considered to be a ``Roper''-like radial excitation. Next is the prominent $\Lambda_c(2625)$ which extends
beyond the top of the plot, and then the
$\Lambda_c(2593)$ peak which is distorted by its proximity to the $\Sigma_c(2455)\pi$ threshold. 

\begin{figure}[h]
\centering
\includegraphics[width=80mm]{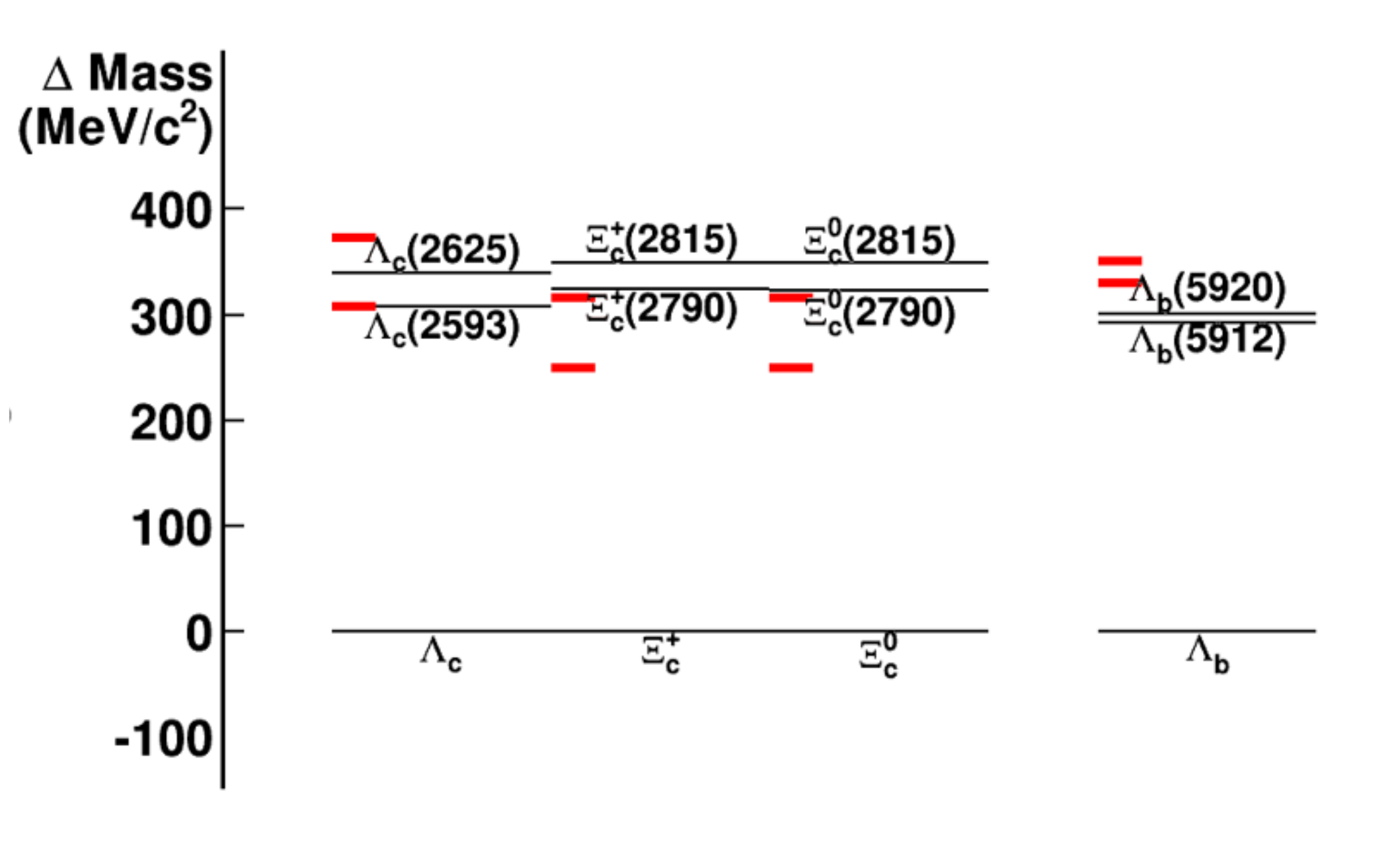}
\caption{The excitation energy (i.e. mass above ground state) of the eight heavy-quark baryons identified as $\lambda = 1$
excitations. The red lines show the rough threshold for preferred decays of the states. Data from the PDG~\cite{PDG}.} 
\label{fig:Masses}
\end{figure}

The identification of the quark structure of 
the $\Lambda_c(2593)$ was made on discovery~\cite{C2595} as being the $J^P=\frac{1}{2}^-$
lowest-mass (``$\lambda$'') orbital excitation. Based on this identification, together with that of the $\Lambda_c(2625)$
as its $J^P=\frac{3}{2}^-$ partner, the scale was set for the excitation energy ($\approx 300\ {\rm MeV}/c^2$)
for a $\lambda\ =\ 1$ orbital excitation and the splitting between the states of $\approx 30\ {\rm MeV}/c^2$. 
According to the standard potential model, the first should be roughly independent of heavy-quark mass, and the
latter approximately inversely proportional to the heavy-quark mass. Thus, this model thus predicted the masses
(as well as the properties) of more than six other states; the $\Xi_c(2790)$ and $\Xi_c(2815)$ iso-doublets, 
and the $\Lambda_b(5912)$ and $\Lambda_b(5920)$ 
have all been subsequently observed.
Figure~\ref{fig:Masses} shows how the expected pattern of masses has been confirmed. 

Despite the success of these predictions, recent papers have conjectured that the $\Lambda_c(2593)$ can be explained
as a molecular state~\cite{MS} or a dynamically generated meson-baryon state~\cite{DGS}. 
The former is motivated by the fact that its mass is very
close to the sum of the $\Sigma_c(2455)$ and  $\pi$ masses. However, it can be seen in Fig.~\ref{fig:Masses} that of the eight states, 
four have masses above the kinematic threshold of their preferred decays, three have masses
below (which keeps the states very narrow) - it is hardly
surprising that one has a mass that coincides, within the fuzziness caused by the isospin mass-splitting of the transition pions, with its particular threshold.
My questions are, if the $\Lambda_c(2593)$ is not the $\lambda = 1$ orbital excitation, how do you explain that the model correctly
predicted the other six states? 
Secondly, what parameter of the $\Lambda_c(2593)$ should we be measuring in order to discriminate between 
the models? The resonance shape 
is clearly complicated by the threshold, but as yet there is nothing observed inconsistent with the quark 
content being dominated by the simplest model.
We rate scientific theories by the validity of their predictions, 
and the (heavy quark)/(light di-quark) model for heavy baryons has served
us very well in this respect. Of course, in any conversation on the inner structure of baryons,
we must always be aware that no one simple model will give the complete story.

\section{$\Xi_c$ Isospin splitting}

There has been considerable progress in recent years on the $\Xi_c$ spectrum. There are six iso-doublets with well-measured masses, plus several
other states not as well defined. When tabulating the masses of the six (Tab.~\ref{tab:isospin}), 
it appears that the isospin mass splitting divides into two groups, one
with around 3.5 ${\rm MeV}/c^2$, and one with less than 1 ${\rm MeV}/c^2$. 
The first group all have a spin-0 di-quark, and the latter
group have a spin-1 di-quark. My question, is this a rule that can then be applied to help 
identify higher mass states that are being discovered?

\begin{table}[h]
\begin{center}
\caption{Isospin mass splitting in the $\Xi_c$ iso-doublets. Data taken from~\cite{PDG} and~\cite{Yelton}.}
\begin{tabular}{c|c}
\hline
Particle & $M(\Xi_c^+)-M(\Xi_c^0)$ ${\rm MeV}/c^2$ \\
\hline 
$\Xi_c$ & $-3.3 \pm 0.4$ \\
\hline 
$\Xi_c(2645)$ & $-0.9 \pm 0.5$ \\
\hline 
$\Xi_c(2815)$ & $-3.5 \pm 0.5$ \\
\hline 
$\Xi_c(2980)$ & $-4.8\pm0.5$ \\
\hline 
$\Xi_c^{\prime}$ & $-0.8\pm 0.5$ \\
\hline 
$\Xi_c(2790)$ & $-3.3\pm 0.6$ \\
\hline
\end{tabular}
\label{tab:isospin}
\end{center}
\end{table}

\section{$\Omega_c^0$ Decays}

Of the four weakly decaying charmed baryons, the least is known about the $\Omega_c$ (css). 
The general knowledge of its decays seemed uncontroversial until recently. The lifetime of the $\Omega_c^0$ had been measured by 
three experiments to be very short, and this agreed with early ideas on the expected lifetime hierarchy of the four 
lifetimes~\cite{OMT}.
However, of the three experiments measuring the lifetimes~\cite{WA,687,FOCUS}, one had signals in various modes 
but never managed to publish an $\Omega_c$ mass~\cite{WA}, and another
used a decay mode that has never been seen by other experiments~\cite{687}, despite a search~\cite{OC}. 
I note that if you over-estimate your signal yields you tend 
to produce a short lifetime measurement. Recently LHCb found a much longer lifetime~\cite{LHCOl}, 
showing that the knowledge of the $\Omega_c^0$ weak decays
was less definitive than had been thought. (I note that one paper~\cite{CHENG} 
brought attention to this problem before the experimental measurement).
Information on the $\Omega_c^0$ branching fractions is comparatively scarce, but one aspect that stands out is 
that the branching ratio $B(\Omega_c^0\to\Omega^-\pi^+\pi^-\pi^+)/B(\Omega_c^0\to\Omega^-\pi^+)$ is much less 
than the analagous ratios for the other 
charmed baryons (Tab.~\ref{tab:Omegac}). In the most simple decay models, these should all be similar - it easy to see how factors of two or three can arise 
from differences
in phase-space etc. but the difference of a factor of ten seems very large. There is a similar trend 
(though not as marked) with the decay into $\Omega^-\pi^+\pi^0$.
My question is whether these branching fractions tell us anything basic about
the differences in the decays of the weakly decaying charmed baryons.

\begin{table*}[t]
\begin{center}
\caption{Branching ratios comparison for charmed baryons. Data taken from~\cite{PDG}.}
\begin{tabular}{c|c|c|c}
\hline
$\Omega_c^0$ & $\Lambda_c^+$  & $\Xi_c^0$ & $\Xi_c^+ $\\
\hline 
$\Omega^-\pi^+\pi^0/\Omega^-\pi^+$&$\Lambda\pi^+\pi^0/\Lambda\pi^+$ & $\Xi^-\pi^+\pi^0/\Xi^-\pi^+$ & $\Xi^0\pi^+\pi^0/\Xi^0\pi^+$  \\
\hline 
$1.97\pm0.17 $  &  $5.46\pm0.42 $ & Not measured & $4.2 \pm 1.5$ \\
\hline 
$\Omega^-\pi^+\pi^-\pi^+/\Omega^-\pi^+$&$\Lambda\pi^+\pi^-\pi^+/\Lambda\pi^+$ & 
$\Xi^-\pi^+\pi^-\pi^+/\Xi^-\pi^+$ & $\Xi^0\pi^+\pi^-\pi^+/\Xi^0\pi^+$  \\
\hline 
$0.29\pm0.04 $  &  $2.84\pm0.34 $ & $3.3\pm0.5$ & $3.1 \pm 1.0 $\\
\hline
\end{tabular}
\label{tab:Omegac}
\end{center}
\end{table*}

\section{Excited $\Omega_c^0$ Spectroscopy}

This is one subject where, rather than too few, there are too many answers. The LHCb~\cite{OMCS} discovery of five narrow states brought an immediate flood 
of explanations. The most obvious (though naive) explanation is that the first $\lambda = 1$ excitation states for the $\Omega_c^0$ will in fact be
a quintuplet of states, and given that higher spin usually indicates higher mass, the states can be labeled as 
$
J^P=(\frac{1}{2}^-,
\frac{1}{2}^-,
\frac{3}{2}^-,
\frac{3}{2}^-,
\frac{5}{2}^-)
$
The Belle analysis~\cite{BOMCS} showed four of the states, but the last one is missing. This did not strike people as surprising given that the fifth 
(i.e. highest mass) of the LHCb quintuplet was their smallest peak. However, in $e^+e^-$ continuum production, the production rates of states within
a family occurs via the ``$2J+1$'' rule. This has been verified in a variety of systems~\cite{NII}. Even considering mass suppression, and the possibility
of decays to $\Xi_c^{\prime}$, it would seem surprising that the highest spin state would not be seen by Belle. The plot thickens with the 
(low statistics) search for these states by LHCb~\cite{LBD} in $\Lambda_b$ decays, which again shows the fifth state missing. My question is this, when
considering the possible assignment of states, why do people not take into account the production ratios? For $e^+e^-$ the ratio between the 
different members of a group (for instance, the $\lambda = 1$ quintuplet) is clear, but no theorist has mentioned them in their attempts to explain the data.
In LHCb's first paper, 
the situation is more confusing as different production mechanisms are mixed together, and in their second they had no 
theoretical prediction of the 
production of the various states in $\Lambda_b$ decays. 
Surely these ratios could have been predicted in advance of the experiment.

\section{Excited $\Sigma_c$ Spectroscopy}  
 
It was way back in 2004 that Belle reported the discovery~\cite{SCSBe} of a new $\Sigma_c$ resonance decaying into $\Lambda_c^+\pi$. 
The natural explanation is that
it is an orbital excitation of the $\Sigma_c$. However, like the $\Omega_c^0$, the $\Sigma_c$ should have a family of a quintuplet of $\lambda = 1$ low-mass
states, so why is
there only one peak?
It is true that some of the five might be too wide to be seen, but I do not believe that any model predicts one to be narrow and four 
to be wide. This certainly indicates that the peak seen may be due to overlapping states. 
Very similar results are shown in 
an (unpublished) analysis using BaBar data~\cite{Ahmed}. I note that BaBar~\cite{SCSB} also see a peak in the same mode, 
but at a different mass, in $B$ decays. 
To disentangle these observations what we need is not just predictions of the masses and widths
of states in this range, 
but also estimates of the relative production rates, particularly in $B$ decays. 
I also cannot help asking the question of why LHCb have not shown
information on this mass spectrum.

\section{Summary}

I have highlighted several questions which I believe could be fruitful lines of research for theorists. They are just a few of many I have! The charmed 
baryon sector is already very rich in information, and the next few years we will see more results,
 so there are many opportunities to make
predictions that can be directly tested.

\begin{acknowledgments}
This work is supported in part be DOE contract DE-SC0009824.
\end{acknowledgments}

\bigskip 

\end{document}